\documentclass[11pt,a4paper]{article}
\usepackage[noadjust]{cite}

\usepackage[margin=2.8cm,bottom=3.5cm]{geometry}

\usepackage{amsmath, amssymb}
\usepackage{color}
\pdfoutput=1 

\usepackage{graphicx} 
\usepackage{subcaption}
\usepackage[T1]{fontenc}

\title{Fermion transfer in the $\phi^4$ model with a half-BPS preserving impurity}

\author{
        Jo\~ao G. F. Campos \\
        Departamento de F\'isica, Universidade Federal de Pernambuco,\\
        Av. Prof. Moraes Rego, 1235, Recife - PE - 50670-901, Brazil\\
        jgfc@df.ufpe.br
            \and
        Azadeh Mohammadi\\
        Departamento de F\'isica, Universidade Federal de Pernambuco,\\
        Av. Prof. Moraes Rego, 1235, Recife - PE - 50670-901, Brazil\\
        azadeh.mohammadi@df.ufpe.br
}

\begin{document} 
\maketitle

\begin{abstract}

We study a fermion field coupled to a scalar via a Yukawa term. The scalar field is the $\phi^4$ model with an impurity that preserves half of the BPS property. We analyze the spectrum of the defects of the model and collisions between them both close to the BPS regime and not. As the fermion binds to these defects, it may be transferred from one to the other, which we quantify via overlaps, known as Bogoliubov coefficients. BPS collisions are less likely to transfer the fermion between defects and can be adiabatic for non-relativistic velocities, especially for small coupling constants. Moreover, closer to the BPS limit only a small fraction of the fermion number is radiated away. In contrast, non-BPS collisions lead to more radiation in the fermion field and excitation of the fermion to higher bound states, and the result is more sensitive to the parameters.
\end{abstract}

\section{Introduction}

Interactions of the fermion field with solitons have been subject to intense research since the pioneering work of Jackiw and Rebbi \cite{jackiw1976solitons}. Interestingly, they found that solitons may have an associated half-integer fermion number whenever there exists a bound zero-energy solution for the fermion field. Later, a series of works has shown that a soliton can have any fractional fermion number \cite{goldstone1981fractional,niemi1986fermion,alonso2019soliton}.
In particular, Jackiw and Rebbi investigated a model where the fermion is coupled with a $\phi^4$ kink via Yukawa coupling, ignoring the back-reaction. This model can be solved analytically and has a well-known set of bound and scattering states as shown for instance in
\cite{chu2008fermions,charmchi2014complete} considering a massless fermion field and in
\cite{charmchi2014massive} a massive one. Since then other kink-fermion systems have been studied. For instance, it is possible to compute the Casimir energy of the fermion field when the fermion is chirally coupled with a prescribed scalar field
\cite{shahkarami2011casimir,gousheh2013casimir,
      gousheh2014investigation}. Another example is the computation of the energy and eigenfunctions of fermion bound states where the background scalar field is a modified sine-Gordon or a modified $\phi^4$ 
\cite{bazeia2017fermionic,bazeia2019fermion}. Moreover, kink-fermion interactions arise naturally in supersymmetric systems \cite{charmchi2014one}. In higher dimensions, fermions have been studied in the background of vortices 
\cite{jackiw1981zero}, chiral fields 
\cite{kahana1984soliton} and skyrmions 
\cite{hiller1986solutions}, for example.

This problem becomes more interesting when one considers a kink-antikink pair, instead of just a kink, as the background. This was done in 
\cite{chu2008fermions} for the $\phi^4$ model where the authors computed the energy spectrum and eigenstates of a fermion in such background. They showed that, as the distance of the kink-antikink increases, the fermion bound states and energy approach the ones of a single kink, as expected. This result was repeated in 
\cite{brihaye2008remarks}, now considering sine-Gordon kink-antikink pair instead of the $\phi^4$. There, snapshots of the exact solution of a kink-antikink collision were considered as the background field, however, without any reference to the problem of bound states after the collision. The issue here is that there are no bound states after the collision. This problem arises in many cases where the kink is not centered around the origin. In 
\cite{brihaye2008remarks} this problem was circumvented by shifting the sine-Gordon kink to center it around the origin, however, after the collision, this is not the case anymore and there is no bound state. Therefore, it is hard to find models where we can discuss fermion bound states in kinks collision backgrounds that go beyond the $\phi^4$ model.

There is a more intriguing problem than computing fermion bound states for a kink-antikink background which is the exchange of fermions between the kinks or the fermions transfer between fermion bound states during a kink and antikink collision, as done in 
\cite{gibbons2007fermions,saffin2007particle}. There, the background scalar field is not fixed anymore and evolves dynamically. During the collision, the fermion is affected by the scalar field and can stay on the kink, be transferred from the kink to the antikink or radiate. This is the type of analysis that we focus on in the present paper. In \cite{gibbons2007fermions,saffin2007particle}, this analysis was motivated by previous works investigating the possibility that higher-dimensional universes can behave like a four-dimensional one if particles are bound to a brane that localizes them in the extra dimensions  
\cite{rubakov1983we,koley2005scalar,
      randjbar2000fermion,melfo2006fermion}. The authors in 
\cite{gibbons2007fermions,saffin2007particle} tried to understand the fate of fermions when such branes collide and it is interesting to have this interpretation in mind. 

It is worth pointing out that it is possible to add another ingredient to the problem: the back-reaction of the fermion on the soliton. It has been shown that a prescribed soliton is a good approximation for small coupling constants and considering the back-reaction can create bound soliton-antisoliton pairs and also can mediate interactions between the solitons \cite{shahkarami2011exact,amado2017coupled,klimashonok2019fermions,perapechka2018soliton,
perapechka2020kinks,perapechka2019fermion}. Here, we study a fermion field coupled to a scalar field via the Yukawa interaction in (1+1) dimensions as in 
\cite{amado2017coupled}. However, as in most works cited above, we consider the scalar field as a background, even for larger values of the coupling constant, because it greatly simplifies the problem, allows some analytical treatment and a more direct comparison to the works mentioned before.

As the scalar field is evolving dynamically in our study, it is important to highlight some relevant works involving kink-antikink collisions. One of the pioneering ones was done by Sugiyama \cite{sugiyama1979kink} where the author estimated the critical velocities in kink-antikink collisions using a collective coordinate approach. A few years laters, Campbell et al. 
\cite{campbell1983resonance} did a precise numerical computation of the pattern of resonance windows. Remarkably, the authors showed that while a kink and an antikink annihilate for small relative velocities and reflect for high relative velocities, there are intermediate velocities where they collide multiple times before separating. Furthermore, they gave an approximate explanation for the resonance windows phenomena as an energy exchange mechanism between the kink translational and vibrational energy.

More recent works in kink-antikink interactions include $\phi^4$ model and modifications 
\cite{takyi2016collective,bazeia2018scattering, dorey2018resonant}; interaction of kinks in higher-order models such as $\phi^6$ and $\phi^8$ 
\cite{takyi2016collective,dorey2011kink,gani2014kink,gani2015kink}; coupled two-component kinks 
\cite{alonso2018reflection, halavanau2012resonance}; models with power-law asymptotics 
\cite{gomes2012highly,belendryasova2019scattering,christov2019kink}; and others 
\cite{simas2017degenerate,marjaneh2017multi}. It is a rich field of research with many interesting works and novel results. More recently some attention has been directed towards the $\phi^4$ model with a half-BPS preserving impurity 
\cite{adam2019phi,adam2019spectral}. In this model, the impurity is a term in the Lagrangian that breaks translational invariance in such a way that the model still admits one BPS solution. The model admits topological and nontopological defects consisting of kinks, antikinks and lumps. During the collisions, some of the interactions between the defects are BPS preserving versus the others which are not. Similar half-BPS preserving models with exactly solvable BPS sector were also considered in \cite{adam2019solvable}. Supersymmetric extensions of these models, where the scalar field naturally couples to a fermion field, were studied in \cite{adam2019bps}.

Here, we take the solutions of the $\phi^4$ model with the half-BPS preserving impurity as a background, coupled to a fermion field similar to \cite{adam2019bps}, although non-supersymmetrically. For a specific range of parameters, the model gives rise to fermion bound states for the fermion interacting with the scalar field configurations. We study fermion transfer where the background is a collision between the defects of this model, with both BPS and non-BPS interactions.

In section \ref{model}, we present the $\phi^4$ model with a half-BPS preserving impurity, interacting with a fermion field via a Yukawa coupling. In section \ref{Results}, we study the time evolution and transfer of the fermion field during a collision between different components in the scalar field. Finally in section \ref{conclusion}, we discuss and summarize our conclusions.

\section{Model}
\label{model}

\subsection{Lagrangian and Euler-Lagrange equations}

We study a model given by the following Lagrangian in $1+1$ dimensions, which can be organized into three types of terms
\begin{equation}
\mathcal{L}=\mathcal{L}_{scalar}+\mathcal{L}_{fermion}+\mathcal{L}_{int}.
\end{equation}
The scalar Lagrangian is the soliton-impurity model studied in 
\cite{adam2019phi}
\begin{equation}
\mathcal{L}_{scalar}=\frac{1}{2}\phi_t^2-\frac{1}{2}\phi_x^2-U(\phi)-2\sigma\sqrt{U(\phi)}-\sqrt{2}\sigma\phi_x-\sigma^2,
\end{equation}
that differs from the typical scalar field theories due to the $\sigma$ term, which describes a half-BPS preserving impurity. The parameter $U(\phi)$ is the scalar potential term depending on the scalar field $\phi(x,t)$. The fermion Lagrangian is given by 
\begin{equation}
\mathcal{L}_{fermion}=i\bar{\psi}\gamma^\mu\partial_\mu\psi,
\end{equation}
and we consider a Yukawa interaction
\begin{equation}
\mathcal{L}_{int}=-g\phi\bar{\psi}\psi.
\end{equation}

The scalar Lagrangian demands some deeper discussion. Following 
\cite{adam2019phi}, we choose the potential as in the $\phi^4$ theory
\begin{equation}
U(\phi)=\frac{1}{2}(1-\phi^2)^2.
\end{equation}
The $\sigma$ terms are added to the Lagrangian such that the system still has one BPS solution resulting from
\begin{equation}
\label{BPS}
\phi_x+\sqrt{2}\sigma+(1-\phi^2)=0.
\end{equation}
This should be compared with the $\phi^4$ model, where two BPS solutions exist, associated with each of the two topological sectors (kink and antikink), instead. The BPS solution in Eq.~\ref{BPS} corresponds to the antikink solution of the $\phi^4$ model. Hence, to find the kink solution we solve the full second-order field equation. To simplify the kink solution, $\sigma$ is chosen such that the $\phi^4$ kink at the origin is still a solution. This leads to \cite{adam2019phi}
\begin{equation}
\label{sigma}
\sigma=\frac{\lambda}{\cosh^2(x)},
\end{equation}
where $\lambda$ is a constant in the range $\lambda>-\sqrt{2}$. In other words, $\phi_{K_0}(x)=\tanh(x)$ solves the field equations when $\sigma$ is given by Eq.~\ref{sigma}. However, $\phi_K(x;x_0)=\tanh(x-x_0)$ does not solve the field equations for this choice of $\sigma$, if $x_0\neq0$. For more details regarding the $\phi^4$ scalar field with this half-BPS preserving impurity see reference 
\cite{adam2019phi}.

The Euler-Lagrange equations for this model are
\begin{align}
\label{phiEL}
\phi_{tt}-\phi_{xx}+2(\phi^2-1)\phi&=\frac{2\sqrt{2}\lambda}{\cosh^2(x)}(\phi-\tanh(x))\\
\label{psiEL}
i\gamma^\mu\partial_\mu\psi-g\phi\psi&=0.
\end{align}
In Eq.~\ref{phiEL} we ignored the term proportional to $g\bar{\psi}\psi$, meaning that we disregarded the back-reaction of the fermion on the scalar field. To solve Eq.~\ref{psiEL}, let us choose a representation for the gamma matrices. We choose the complex representation $\gamma^0=-\sigma^2$, $\gamma^1=i\sigma^3$. In this representation, the fermion field can be split into two decoupled Majorana fields
\begin{equation}
\psi=\psi^M_1+i\psi^M_2.
\end{equation}
Each of these fields has two real components. We ignore the second Majorana field $\psi^M_2$ because it has identical equations to $\psi^M_1$. Writing $\psi^M_1=(\psi_1,\psi_2)^T$, the Euler-Lagrange equation becomes
\begin{align}
\label{fermion1}
\partial_t\psi_1&=-\partial_x\psi_2+g\phi\psi_2,\\
\label{fermion2}
\partial_t\psi_2&=-\partial_x\psi_1-g\phi\psi_1.
\end{align}

\subsection{Scalar field solutions}
\label{scalar}

The solutions discussed in this section were originally found in \cite{adam2019phi}. Let us consider static solutions of the scalar field first. The first interesting static solution is the kink-on-impurity $K_0$ given by $\phi_{K_0}=\tanh(x)$. The subscript $0$ indicates that it is bound to the impurity. As discussed before, the model was constructed such that this is still a solution of the field equations as can be seen in Eq.~\ref{phiEL}. 

\begin{figure}[tbp]
\centering
     \includegraphics[width=0.5\linewidth]{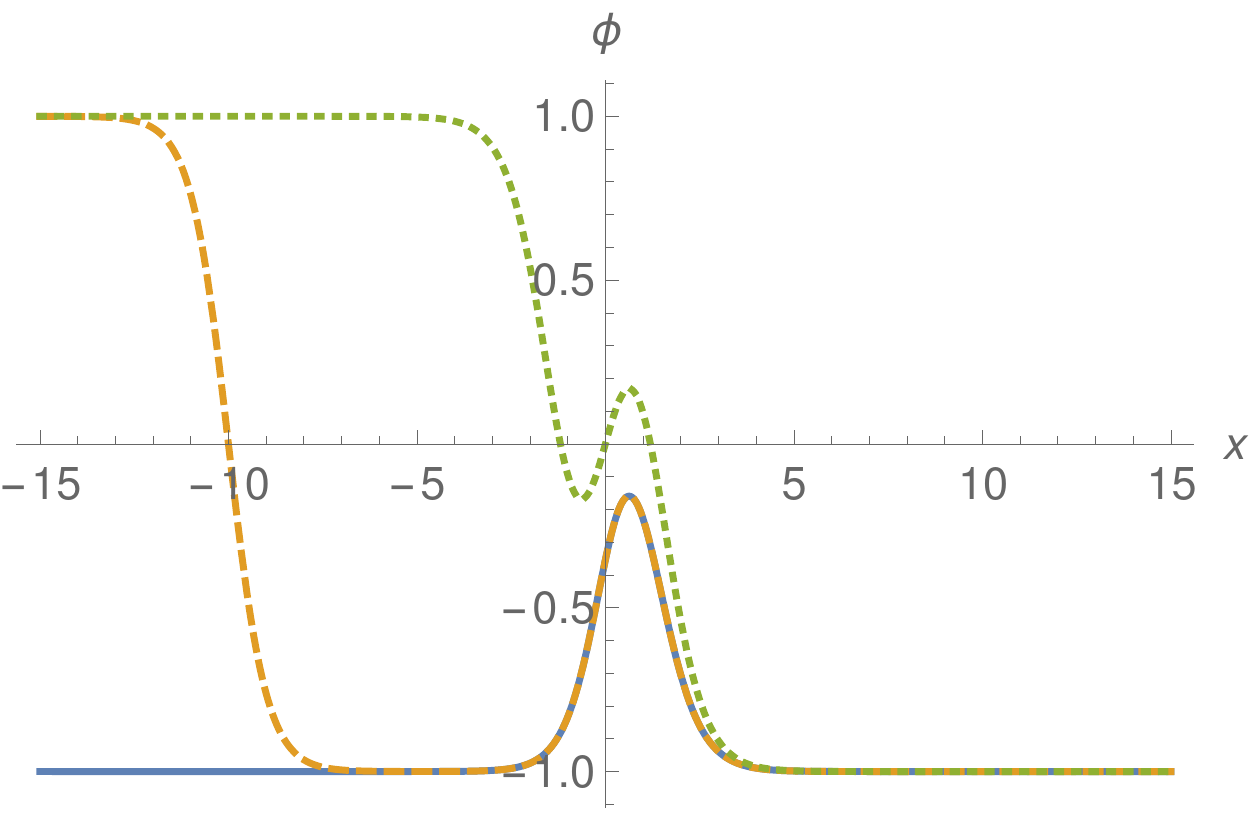}
       \caption{BPS solutions of the scalar field. The full line is the lump solution, The dashed line is an antikink to the left and a lump and the dotted line is the antikink-on-impurity solution. This figure is a reproduction of results in \cite{adam2019phi}.}
       \label{antikink-lump}
\end{figure}

The second interesting static solution is the antikink. As the BPS property is preserved for antikink solutions, they consist of a family of solutions related by a generalized translational symmetry as shown in \cite{adam2019phi}. These solutions $\phi_{AL}(x;x_0)$ are the full BPS antikink solutions. They can be parameterized by a coordinate $x_0$ and usually consist of a $\phi^4$ antikink $A$ and a lump $L$. They can be found numerically integrating Eq.~\ref{BPS} with different initial conditions. We choose the parameter $x_0$ such that the initial condition is $\phi_{AL}(x_0;x_0)=0$. We fix $\lambda=-1.0$ for the reason which will be discussed shortly. The antikink solution symmetric around the origin is called antikink-on-impurity $A_0$. It is given by $\phi_{A_0}(x)\equiv\phi_{AL}(x;0)$ and is shown in Fig.~\ref{antikink-lump} (dotted line). This solution resembles a kink at the origin surrounded by two symmetric antikinks. As we translate the antikink from the origin the solution becomes a translated antikink $A$ and a lump which is near the origin, where the impurity is. This is shown in Fig.~\ref{antikink-lump} (dashed line). Finally, in the limit that the antikink is translated to plus or minus infinity the solution consists of only a lump $\phi_{L^\pm}(x)=\phi_{AL}(x;\pm\infty)$ as shown in Fig.~\ref{BPS} (solid line) for $L^-$. The two lump solutions $L^\pm$ differ by the property that $\phi_{L^+}(\pm\infty)=1$, while $\phi_{L^-}(\pm\infty)=-1$. 

Next, we would like to build approximate composite solutions using the additive ansatz. This can be built using a solution of the complete field equations (such as $\phi_{K_0}$, $\phi_{A_0}$ and $\phi_{L^\pm}$) and a $\phi^4$ solution far from the origin. For instance
\begin{equation}
\phi(x)=\phi_{K_0}(x)+\phi_A(x;x_0)+1,\label{additive ansatz}
\end{equation}
where $\phi_A=-\tanh(x-x_0)$ is the solution of the $\phi^4$ antikink $A$. This is an approximate solution only for $x_0\ll-1$. If one replaces $+1$ by $-1$ in the above equation, the condition changes to $x_0\gg1$. It is easy to see that Eq. \ref{additive ansatz} is a solution of the field equations except for an exponentially decreasing overlap. The same is true if we add a boosted antikink $\phi_A(x,t;x_0,v)=-\tanh(\gamma(x-vt-x_0))$ to the kink-on-impurity
\begin{equation}
\label{inicond1}
\phi(x,t)=\phi_{K_0}(x)+\phi_A(x,t;x_0,v)+1,
\end{equation}
where again $x_0\ll-1$. We will discuss the evolution of this solution in the following sections. 

Using a similar reasoning, we can approximate the BPS solution $\phi_{LA}(x,x_0)$ for $x_0\ll-1$ using the additive ansatz considering a static antikink $\phi_A(x;x_0)$
\begin{equation}
\phi_{LA}(x;x_0)\simeq\phi_{L^-}(x)+\phi_A(x;x_0)+1. \label{additive ansatz2}
\end{equation}
or the boosted one $\phi_A(x,t;x_0,v)$, where the solution is close to the BPS regime for small $v$.
Moreover, it is also possible to build solutions with the $\phi^4$ kink $K$ the same way. In the following sections, we will consider the evolution of the aforementioned solutions treating the scalar field classically.

\subsection{Fermion bound states}

\label{fermion-bs}

Now, let us study the fermion field in the presence of static or boosted solutions of the scalar field discussed in the previous section. The bound states are found by solving Eqs. \ref{fermion1} and \ref{fermion2}. First we use the ansatz $\psi_1=\eta_+\cos(\omega t-\theta)$ and $\psi_2=\eta_-\sin(\omega t-\theta)$. After substituting one equation into the other, this gives two decoupled equations for $\eta_\pm$
\begin{equation}
\label{schrodinger}
-\partial^2_x\eta_\pm+g(g\phi^2\mp\partial_x\phi)\eta_\pm=\omega^2\eta_\pm,
\end{equation}
which is a Schr\"{o}dinger-like equation with the potential $V_\pm=g(g\phi^2\mp\partial_x\phi)$. These equations have well-known solutions for the $\phi^4$ kink and antikink as shown in 
\cite{chu2008fermions,charmchi2014complete}. For instance, the fermion zero mode of the $\phi^4$ kink centered at $x_0$ is given by
\begin{equation}
\psi_1=\mathcal{N}\cosh^{-g}(x-x_0),\quad\psi_2=0,
\end{equation}
where $\mathcal{N}$ is the normalization constant. The full discrete spectrum is given by
\begin{equation}
\omega_n=\sqrt{n(2g-n)},\quad0\leq n<g,\quad n\in Z^+.
\end{equation} 
The fermion zero mode always exists for this model and the first excited state appears for $g\geq 1$. Therefore, we set $g\geq 1$ to include the first excited state of the kink in our analysis. 

The solutions can be boosted in the standard way. We set $x^\prime=\gamma(x-vt)$ and $t^\prime=\gamma(t-vx)$ together with
\begin{align}
\psi^\prime_1(x,t)&=\cosh(\chi/2)\psi_1(x^\prime,t^\prime)+\sinh(\chi/2)\psi_2(x^\prime,t^\prime),\\
\psi^\prime_2(x,t)&=\sinh(\chi/2)\psi_1(x^\prime,t^\prime)+\cosh(\chi/2)\psi_2(x^\prime,t^\prime),
\end{align}
where we defined the rapidity $\chi=\tanh^{-1}v$. For the other static solutions $\phi_{LA}(x;x_0)$, Eq.~\ref{schrodinger} is solved numerically as discussed below. 

\begin{figure}[tbp]
\centering
     \includegraphics[width=0.9\linewidth]{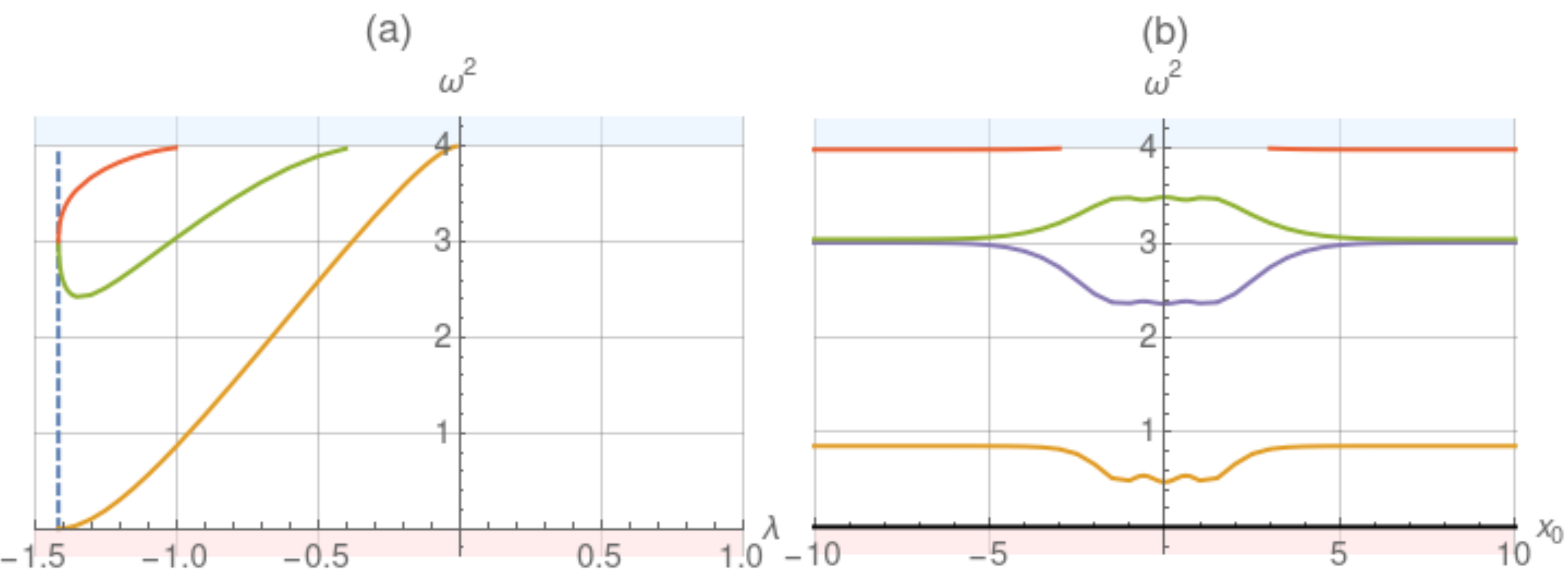}
       \caption{Spectrum of the fermion field coupled to (a) the lump and (b) to the BPS antikink with $\lambda=-1.0$. We set $g=2.0$ (the supersymmetric case). The spectrum is identical to the scalar field spectrum in \cite{adam2019phi}, as expected.}
       \label{spectrum}
\end{figure}

We studied the spectrum of the fermion field bound to the lump with $g\geq 1$ and found that there are only bound states if $\lambda<0$ as shown in Fig.~\ref{spectrum} for $g=2.0$. Therefore, we choose a negative value of $\lambda=-1.0$. Moreover, for this value of $g$, the system coincides with the supersymmetric case and, similar to the discussion in  \cite{adam2019bps}, the spectrum of the fermion bound to the lump is the same as the spectrum of scalar field perturbations around the lump shown in \cite{adam2019phi}. The fermion spectrum in the lump background has no bound states for $\lambda>0$ because $V_\pm$ has no minimum in this case. As we decrease $\lambda$ below zero, a minimum in $V_\pm$ appears together with a fermion bound state. Decreasing further, we approach the limit where the lump becomes a kink-antikink pair giving rise to a fermion zero mode and two degenerate discrete modes with $\omega^2=3$. The depth of the potential $V_\pm$ increases with $g$ and more fermion bound states appear accordingly.

The fermion spectrum bound to BPS antikink can also be computed numerically. It is shown in Fig.~\ref{spectrum}(b) for $\lambda=-1.0$ again in the supersymmetric case $g=2.0$. This can also be compared with the spectrum of the perturbations in the scalar field shown in \cite{adam2019phi}. In the limit $x_0\to\pm\infty$ we see that the fermion spectrum approaches the values of the separate lump and $\phi^4$ antikink, which consists of the zero fermion mode and first excited state bound to the $\phi^4$ antikink and the three discrete fermion states bound to the lump. For $x_0\simeq 0$, the spectrum is slightly deformed. Notice that the highest excited state of the BPS antikink shown in Fig.~\ref{spectrum}(b) disappear in the continuum for small values of $x_0$ similarly to what happens in \cite{adam2019spectral}. This could have interesting effects and be the subject of future investigation. The result of the study will be
reported elsewhere.

\section{Results}
\label{Results}

\subsection{Scalar field collisions}

\begin{figure}[tbp]
\centering
   \begin{subfigure}[b]{0.85\linewidth}
     \includegraphics[width=\linewidth]{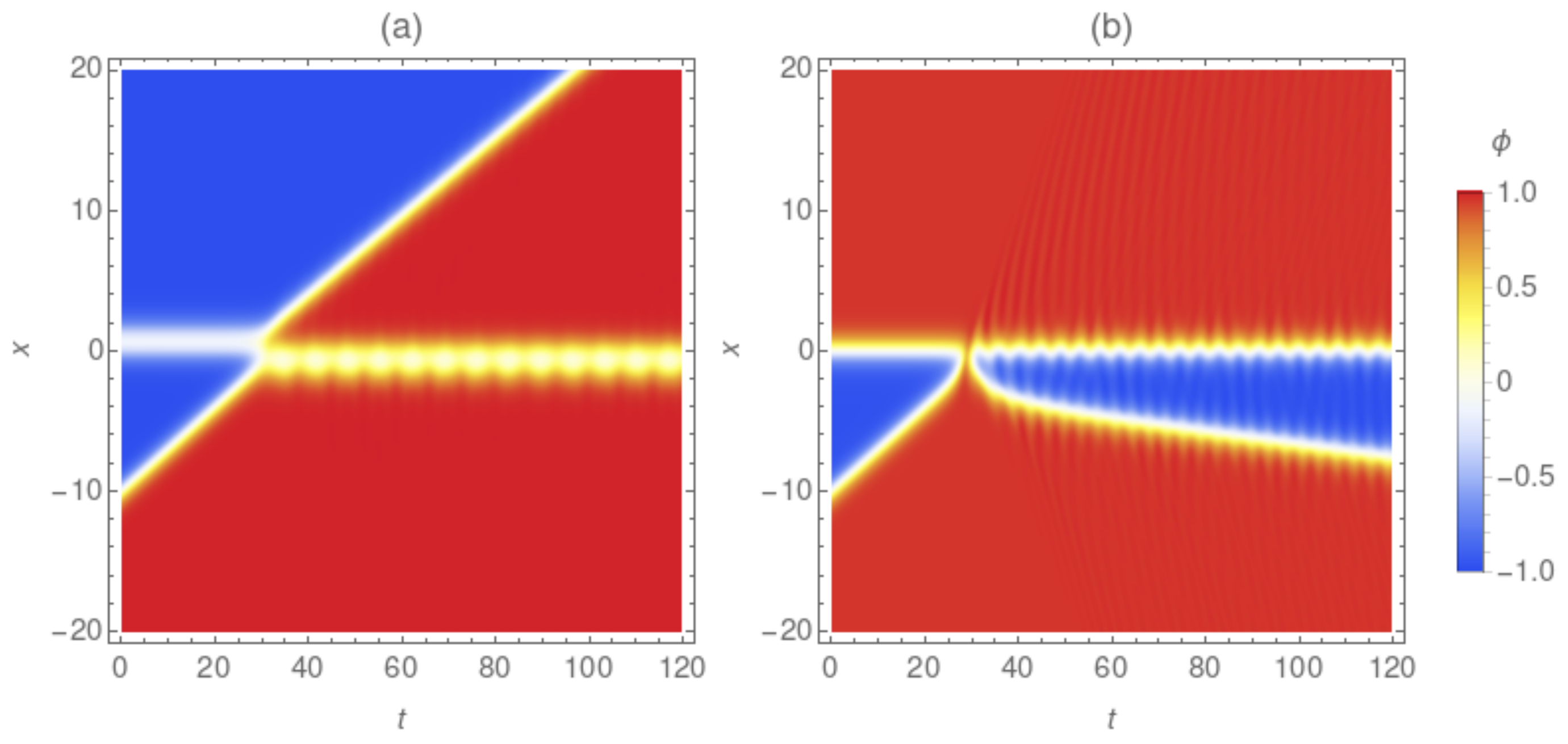}
   \end{subfigure}
   \begin{subfigure}[b]{0.85\linewidth}
     \includegraphics[width=\linewidth]{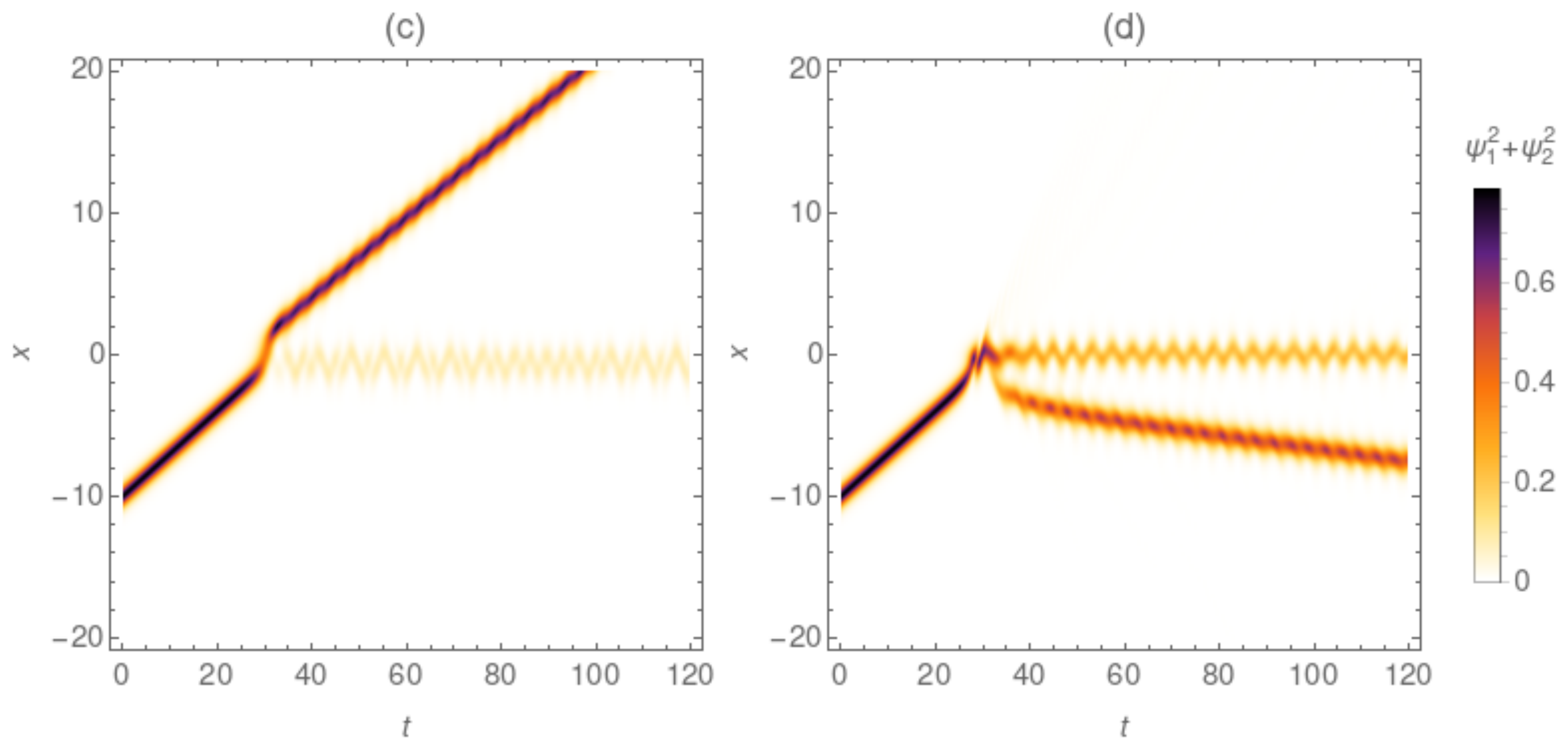}
   \end{subfigure}
       \caption{Upper graphs: Evolution of the scalar field during a collision between (a) an antikink and a lump and (b) an antikink and a kink-on-impurity. These two graphs are the reproduced results in Figs. 14 and 24 in \cite{adam2019phi} with different parameters. Lower graphs: Evolution of the fermion field during the background collision between (c) the antikink and the lump and (d) the antikink and the kink-on-impurity. Parameters are $g=2.0$, $v=0.3$ and $\lambda=-1.0$.}
       \label{phi-and-density}
\end{figure}

Now let us study collisions between defects of the scalar field, as done in \cite{adam2019phi}. It is necessary to repeat the computation here before including the fermion field, however, we will be brief. The details of the numerical integration are given in appendix \ref{ap1}. Here, we consider two types of collisions, one with and one without BPS interactions, among the ones investigated in \cite{adam2019phi}. The first type of collision is between an antikink and a lump. The initial condition for this collision is given by Eq.~\ref{additive ansatz2} with a boosted antikink, which occurs very close to the BPS regime as discussed in section \ref{scalar}. For the value of the parameters used, the kink passes smoothly through the lump which is possible to see in Fig.~\ref{phi-and-density}(a). The process can be written schematically as
\begin{equation}
\label{process1}
A+L^-\to L^++A.
\end{equation}
Notice that after the antikink passes through the lump $L^-$, the lump becomes $L^+$ to adjust with the boundaries.

The next collision we consider is between an antikink and a kink-on-impurity. The initial condition for this collision is given by Eq.~\ref{inicond1}. The kink is tightly bound to the impurity for the chosen values of the parameters. Therefore, the antikink is reflected and the kink remains at the impurity after the collision as shown in Fig.~\ref{phi-and-density}(b). This can be written as
\begin{equation}
\label{process2}
A+K_0\to A+K_0.
\end{equation}
Reflection happens for high velocities only, while for small velocities the kink and the antikink annihilate or resonate. This behavior is reminiscent of the $\phi^4$ theory and we will consider only the values of $v$ where the antikink reflects because otherwise, the system does not have a well defined final state. In the following subsections, we will study the evolution of the fermion field in these two background collisions.

\subsection{Bogoliubov Coefficients}
We treat the fermion field quantum mechanically in contrast with the scalar field which is treated classically. 
Now let us discuss the formalism necessary to study the time evolution of the fermion field in the two scenarios discussed in the previous subsection. 
To do so, we consider a fermion field localized on a defect before the collision in the asymptotic past at $t=0$ and find the fermion field evolution in time via 
the Bogoliubov coefficients $B_{j\to k}$
\begin{equation}
\psi_{in}^j(t)=\sum_kB_{j\to k}(t)\psi_{out}^k(t),
\end{equation}
where the indices $j$ and $k$ specify the type of the defects present initially and after the collision, respectively, with the fermion field bound to them. In the above equation, $\psi_{in}^j(t)$ is the initial fermion bound state evolved in time and initially localized on the defect type $j$ at one of its associated bound states. Time evolution is done integrating the equations of motion. On the other hand, $\psi_{out}^k(t)$ is the final fermion \text{state} bound to the defect type $k$ present after the collision at time $t$. Therefore, the coefficients are given by
\begin{equation}
B_{j\to k}(t)=(\psi_{out}^k(t),\psi_{in}^j(t))\equiv\int(\psi_{out}^k(t))^T\psi_{in}^j(t)dx
\end{equation}
The interpretation of the Bogoliubov coefficient is that $(B_{j\to k})^2$ is the fraction of fermion number transferred to the state $k$ in time $t$, starting from state $j$ in the asymptotic past. This is shown, for example, in \cite{saffin2007particle}, where one can see more details regarding the Bogoliubov coefficients in this context.

\subsection{Adiabatic Evolution}

The numerical techniques employed here to evolve the fermion field are discussed in appendix \ref{ap1}. Let us first discuss the evolution of the scalar and fermion fields in the BPS case for small velocities. When the velocities are small, the scalar field evolves slowly and smoothly from one BPS state to the next and, thus, the evolution is adiabatic. Moreover, if the evolution is truly adiabatic, the fermion field would evolve smoothly from one bound state configuration to the next corresponding configuration as the BPS antikink moves in moduli space.

We will specialize to the case where the fermion field starts at the zero mode of the BPS antikink. A typical collision in the BPS sector is shown in Fig.~\ref{phi-and-density}(a), for the scalar field, and in Fig.~\ref{phi-and-density}(c), for the fermion field considering $v=0.3$. The plot of the fermion field shows the fermion density $n$ defined as
\begin{equation}
n=\psi_1^2+\psi_2^2.
\end{equation} 
The adiabatic limit, however, occurs for $v\lesssim0.1$. The difference between an adiabatic evolution and the nonadiabatic one, the one shown in Fig.~\ref{phi-and-density} for example, is that the passage of the antikink through the lump in the adiabatic evolution is smoother and the vibrational mode of the lump is not excited. Moreover, in this case, there is no fermion density near $x=0$ after the collision, meaning that the fermion field is not transferred from the zero mode to the first excited state, located at the lump.

\begin{figure}[tbp]
\centering
     \includegraphics[width=0.9\linewidth]{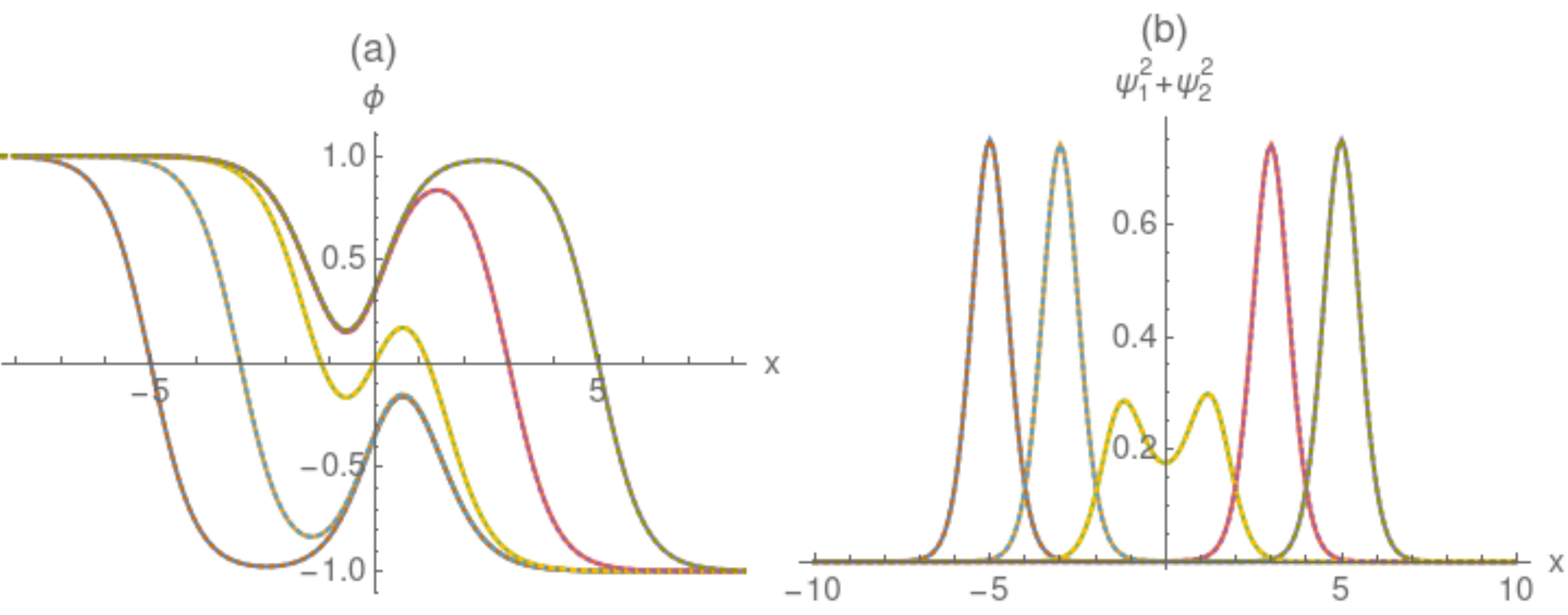}
       \caption{(a) Snapshots (solid) of the scalar field configuration during an adiabatic evolution of an antikink-lump collision with $v=0.02$ and $g=2.0$. Superimposed in the curves we have the static BPS antikink solution (dotted) and the two curves are indistinguishable. (b) Same as before for the fermion density. The dotted curves are now the density of the fermion eigenstates coupled to the corresponding static BPS antikinks.}
       \label{adiabatic}
\end{figure}

To show that the evolution is adiabatic for small velocities we show snapshots of the fields evolution in the BPS sector for $v=0.02$ and $g=2.0$ in Fig.~\ref{adiabatic}. In Fig.~\ref{adiabatic}(a) we have snapshots of the scalar field configuration when it crosses $x_0\simeq-5.0$, $-3.0$, $0.0$, $3.0$ and $5.0$, from left to right. Superimposed in the figure, dotted curves are corresponding to the BPS antikink solution at these positions. The two curves are indistinguishable corroborating the assumption that the evolution is adiabatic. In Fig.~\ref{adiabatic}(b) we show similar plots of snapshots of the fermion density. As before, we superimpose the fermion density of the fermion zero mode bound to the corresponding BPS antikinks and, again, the two curves are indistinguishable. Therefore, the fermion field also evolves adiabatically. This also means that, computing the Bogoliubov coefficient from the initial fermion zero mode of BPS antikink to the same one in the new position gives exactly $1.0$ (within the numerical precision) during the whole evolution.

\subsection{Relativistic Evolution}

Now let us discuss the behavior of the fermion field in relativistic collisions. We initialize the fermion field in the fermion zero mode bound to the boosted $\phi^4$ antikink, denoted by $A$ as before. The result of the fermion field evolution can be used to compute Bogoliubov coefficients of the type $B_{A\to j}$, where $j$ is any defect present in the final state with an attached fermion bound state. The fermion density is plotted in Figs.~\ref{phi-and-density}(c) and (d) for a specific set of parameters. In (c), we observe the evolution of the process in Eq.~\ref{process1} and, in (d), for the process in Eq.~\ref{process2}. In both graphs $n$ is localized around the antikink before the collision, reflecting the initial condition chosen for the fermion field. After the collision, the density is split between the defects and, in general, the split is uneven. Interestingly, most of it is still localized on some defect, instead of on the bulk, similarly to what was found in 
\cite{gibbons2007fermions,saffin2007particle}. The ``amount'' of the fermion field transferred to each defect after the collision can be quantified by the Bogoliubov coefficients and varies with the parameters of the model. Moreover, after each collision, some density may be lost as radiation.

We make the following definitions 
\begin{equation}
B_{A\to A}\equiv\alpha,\quad B_{A\to K}\equiv\beta,\quad B_{A\to AE}\equiv\gamma,\quad B_{A\to KE}\equiv\delta,\quad B_{A\to L}\equiv\xi,
\end{equation}
where $K$ is the fermion zero mode bound to the $\phi^4$ kink, while adding $E$ means we are condidering the first excited fermion bound state instead. Also $L$ denotes the fermion lowest state bound to the lump. Then, we investigate how the Bogoliubov coefficients evolve with time $t$. However, we must be careful with the definition of the coefficients because the defects present before the collision may be different from the defects present after the collision. In particular, for the process in Eq.~\ref{process1}, the lump $L^-$ becomes $L^+$ after the collision. Thus, we define $\xi$ to be the amount of fermion number transferred from $A\to L^-$ before the collision and from $A\to L^+$ after the collision. For the process in Eq.~\ref{process2}, there is no confusion in the definitions because there are a kink(-on-impurity) and an antikink both before and after the collision.

\begin{figure}[tbp]
\centering
     \includegraphics[width=0.95\linewidth]{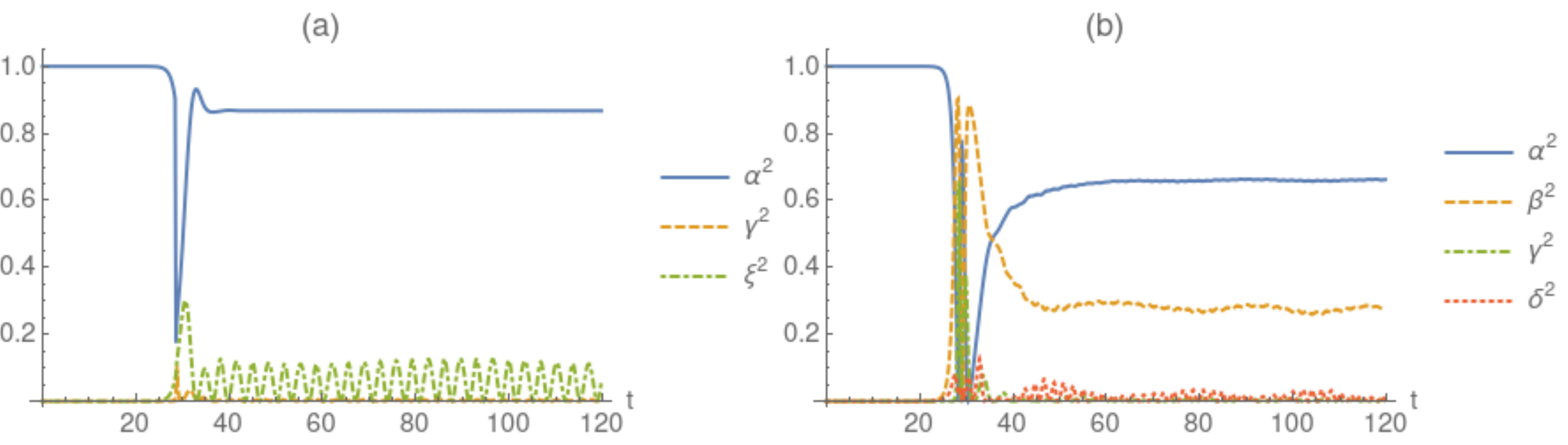}
       \caption{Evolution of the Bogoliubov coefficients as a function of the final time $t_f$. (a) Corresponds to the antikink-lump collision and (b) to the collision between antikink and kink-on-impurity. Parameters are $v=0.3$, $g=2.0$ and $\lambda=-1.0$.}
       \label{bogvst}
\end{figure}

The evolution of the Bogoliubov coefficients with time is shown in Fig.~\ref{bogvst}(a) and (b) for the processes in Eqs.~\ref{process1} and \ref{process2}, respectively. The parameters are the same as in Fig.~\ref{phi-and-density}. We observe that, before the collision, the fermion is completely localized on the antikink, that is, $\alpha^2=1$ and the other coefficients are zero, due to our choice of initial conditions. During the collision, our analysis is not reliable due to the fact that one cannot separate different defects. After the collision, the coefficients rapidly reach a steady state, meaning that the fermion is now bound to the final defects. 

Some points should be noticed. First, the sine (or cosine) dependence in the ansatz (above Eq.~\ref{schrodinger}) of the fermion bound states means that the components of the fermion fields oscillate with phase $\omega t-\theta$. As $\omega\neq 0$ for $\xi$, $\gamma$ and $\delta$ we must be careful when we compute these Bogoliubov coefficients. If the fermion is in one of these states the fermion field $\psi$ will oscillate with phase $\omega t-\theta_0$ for some unknown $\theta_0$. As we do not know the phase $\theta_0$, we project the fermion field at the bound state with a different phase $\omega t-\theta$ to compute the Bogoliubov coefficients, which we fix to an arbitrary constant. This constant will only be equal to $\omega t-\theta_0$ once in a full period. Thus, the coefficients oscillate with time and the amplitude of this oscillation should be taken as the real coefficient. Second, we also see an oscillation in $\beta^2$ in Fig.~\ref{bogvst}(b). This oscillation is accompanied by a negatively correlated oscillation in the amplitude of $\delta^2$. Observing Fig.~\ref{phi-and-density}(b) closely, this can be traced back to the oscillation of $K_0$ that occurs after the collision. This means that $\psi^K$ and $\psi^{KE}$ are not exact bound states of this oscillating $K_0$ and, therefore, there is a transition between the states which is an interesting phenomenon. 
To clarify why this transition occurs, recall from section \ref{fermion-bs} that we know how to compute the Bogoliubov coefficients for two cases: static solutions and their boosts. After the collision, the kinks and lumps can also have the vibrational modes excited, but this effect is usually small and the Bogoliubov coefficients can be computed neglecting this effect still with high accuracy. However, in the collision between an antikink and the kink-on-impurity the final state is neither a static solution nor a boosted one, it is an oscillating kink. The deviation from the static kink solution is not small and cannot be neglected. Luckily, even an oscillating kink has a confining potential and the fermion density stays bound to this kink with the difference that the fermion states bound to the static kink are not the exact bound states of the oscillating kink. Hence, there appears a transition between the states and consequently an oscillation in the Bogoliubov coefficients.

\begin{figure}[tbp]
\centering
       \includegraphics[width=0.9\linewidth]{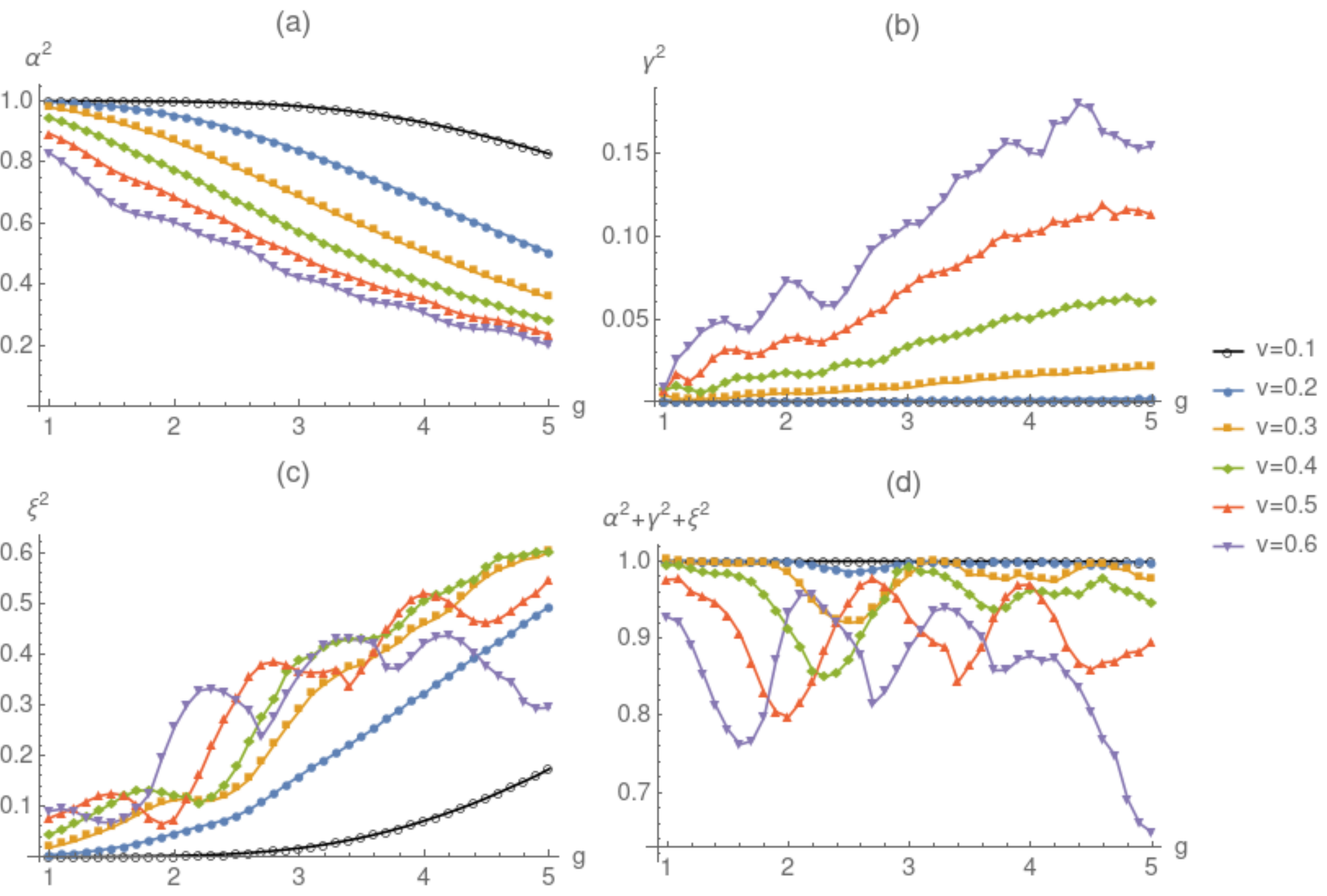}
       \caption{Bogoliubov coefficients versus $g$ for an antikink-lump collision with different values of $v$. We take $\lambda=-1.0$.}
       \label{fermion-antikink-lump}
\end{figure}

\begin{figure}[tbp]
\centering
		\includegraphics[width=0.9\linewidth]{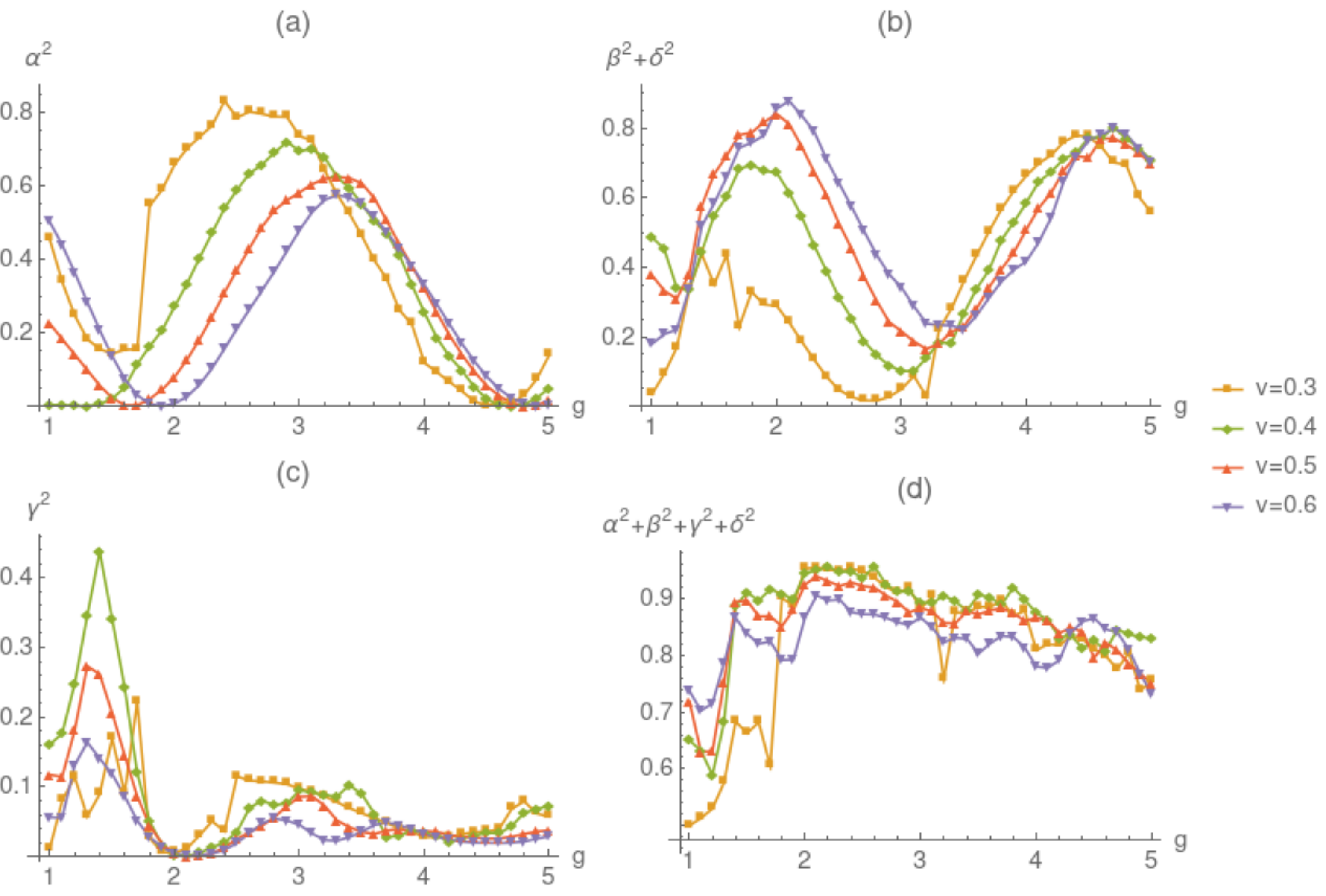}
        \caption{Bogoliubov coefficients versus $g$ for the collision between an antikink and the kink-on-impurity, with different values of $v$. We take $\lambda=-1.0$.}
       \label{fermion-antikink-kinkonimp}
\end{figure}

Now let us investigate the behavior of the final Bogoliubov coefficients as a function of the parameters of the model, $v$ and $g$. These parameters measure, respectively, the velocity of the incoming antikink and the strength of the coupling between the fermion and scalar fields. The results are shown in Figs.~\ref{fermion-antikink-lump} and \ref{fermion-antikink-kinkonimp} for the processes in Eqs.~\ref{process1} and \ref{process2}, respectively.

Consider the antikink-lump collision first. In Fig.~\ref{fermion-antikink-lump}(a) we see the amount of fermion number associated with the zero mode that stays bound to the antikink after the collision, $\alpha^2$. For small $v$, the collision is close to the BPS regime and most fermions stay at this mode. Moreover, in this case the system is closer to the adiabatic limit, where only the fermion zero mode is excited. On the other hand, as we increase $v$, i.e. more distant from the BPS regime, more fermions are transferred to the excited state or the lump. This is quantified by $\gamma^2$ and $\xi^2$ shown in Figs.~\ref{fermion-antikink-lump}(b) and (c). Similarly, if we increase $g$, the fermions are more likely to be affected by the collision and be transferred to the lump even near the BPS regime. Clearly, $\alpha^2$ must be negatively correlated with $\gamma^2$ and $\xi^2$ as shown in the figures. Finally, in Fig.~\ref{fermion-antikink-lump}(d) we plot the sum of the Bogoliubov coefficients in the previous graphs. Table \ref{Bogul-sum}, the left panel, show some values of the sum for example. We find that close to the BPS regime the sum is equal to $1$, meaning that almost all fermions stay at the lowest bound states. Nevertheless, as we increase $v$ more fermions are lost as radiation or transferred to higher excited states, as expected intuitively.  

\begin{table}
\centering
\begin{tabular}{c c c|c c c}
$v$ & $g$ & $\alpha^2+\gamma^2+\xi^2$ & $v$ & $g$ & $\alpha^2+\beta^2+\gamma^2+\delta^2$\\
\hline
0.1&1.0& 1.000 &0.3&1.0& 0.500\\
0.1&2.0& 1.000 &0.3&2.0& 0.953\\
0.1&3.0& 1.000 &0.3&3.0& 0.883\\
0.1&4.0& 1.000 &0.3&4.0& 0.809\\
0.1&5.0& 1.000 &0.3&5.0& 0.754\\
0.2&1.0& 1.000 &0.4&1.0& 0.651\\
0.2&2.0& 1.000 &0.4&2.0& 0.946\\
0.2&3.0& 0.998 &0.4&3.0& 0.892\\
0.2&4.0& 0.995 &0.4&4.0& 0.875\\
0.2&5.0& 0.998 &0.4&5.0& 0.830
\end{tabular}
\caption{The sum of the Bogoliubov coefficients for some values of $g$ and $v$. The left columns correspond to the BPS case and the right columns to the non-BPS case.}
\label{Bogul-sum}
\end{table}

After this analysis, we could compare our results with a non-BPS case from other models such as $\phi^4$. The same analysis we did here was done for the $\phi^4$ model in \cite{gibbons2007fermions,saffin2007particle}. The main difference between the two results is that in the non-BPS case the initial fermion zero mode is much more likely to detach from the antikink and the coefficients are more sensitive to the parameters of the model. Moreover, more fermions are lost as radiation or transferred to higher excited states. Nevertheless, it is also relevant to study a non-BPS case within the same model. Therefore, to complete the analysis, we will now study the Bogoliubov coefficients for the collision between an antikink and the kink-on-impurity in our model. We will show that the results are similar to the ones found for the $\phi^4$ model in  \cite{gibbons2007fermions,saffin2007particle}. 

The final Bogoliubov coefficients for a collision between an antikink and the kink-on-impurity are shown in Fig.~\ref{fermion-antikink-kinkonimp}. As mentioned before, we consider only $v\gtrsim 0.3$ because for smaller values of $v$ the kink and antikink annihilate and we do not have a well defined final state. We see the curve for the coefficient $\alpha^2$ in Fig.~\ref{fermion-antikink-kinkonimp}(a) which shows that for a large interval of the parameters the fermion does not stay at zero mode bound to the antikink in contrast with the BPS case. The behavior is approximately sinusoidal. The curves follow this behavior approximately as argued in \cite{gibbons2007fermions} considering the Dirac equation with an ansatz for the fermion field symmetric concerning $x$ with a time-dependent amplitude and phase interacting with a scalar field approximated by its maximum at the collision during a short time. In this simplified model the authors discarded the bulk fermions and showed that with these approximations a sinusoidal behavior is expected. In Fig.~\ref{fermion-antikink-kinkonimp} the curve for $\alpha^2$ is negatively correlated with the sum $\beta^2+\delta^2$ shown in Fig.~\ref{fermion-antikink-kinkonimp}(b). We have plotted the sum instead of the two individual quantities because, as discussed before, there is a transition between the two states and therefore, the separate quantities are not reliable. It is clear from the curves that we can find very different results varying the parameters of the model in the range considered, meaning that the result is more sensitive to these parameters. The curves (a) and (b) for $v=0.3$ and small values of $g$ show strange behavior such as a jump in $\alpha^2$ near $g\simeq 2$. This can be traced back to the fact also observed in 
\cite{gibbons2007fermions,saffin2007particle} that for small $g$ the fermion bound states are too delocalized compared with the kink size and are more likely to escape when perturbed. 

In Fig.~\ref{fermion-antikink-kinkonimp}(c) we have $\gamma^2$, the fermion excited state bound to the antikink, which usually corresponds to a small fraction of fermion number, but can go as high as $45\%$. Finally, In Fig.~\ref{fermion-antikink-kinkonimp}(d) we have the total probability of the fermions staying in the lowest bound states. Again we show some of the values of the sum in Table \ref{Bogul-sum} for reference. The difference from unity is equal to the amount of fermion number that is transferred to higher excited states or radiated away. We observe that as we increase $v$ the difference from unity becomes larger (except for $g\lesssim 2.0$), as expected to happen when the energy of the system is increased leading to the loss of a larger fraction of fermions in the form of radiation or excitation to higher states, as in the antikink-lump collision. On the other hand, in the antikink-lump collision close to the BPS regime with small $v$ almost no radiation is produced and higher states are not excited.

\section{Conclusion}
\label{conclusion}

The main goal in this work is to compare the fermion transfer between solitons when these solitons collide in BPS and non-BPS cases. In order to do this, we added a fermion field and a Yukawa interaction to a model recently proposed in the literature \cite{adam2019phi} that consists of the $\phi^4$ model with a half-BPS preserving impurity. This model contains different defects that may interact in a BPS or a non-BPS way and it is interesting because it may serve as a guide for higher dimensional soliton interactions where, contrary to the $(1+1)$ dimensional case, there might be multi-soliton BPS solutions. The same is true for our work: it may also serve as a guide for higher-dimensional cases.

We discussed the spectrum of the defects of the model. In particular, we showed that the lump has fermion bound states only for $\lambda<0$ and that the spectrum of the BPS antikink approaches separate spectra of the lump and the $\phi^4$ antikink for large positions in moduli space. As one expects, the spectrum of fermion field is similar to the spectrum of scalar field excitations in the supersymmetric limit. Then, we computed the time evolution of the scalar and fermion fields for two scenarios: a collision between an antikink and a lump and between an antikink and the kink-on-impurity. In both cases, the fermion field is initially bound to the antikink at the zero mode. We found that after the collision, when the defects separate, most of the fermion density is found at the defects and not at the bulk, meaning that the fermion stays bound to the defects even after the collision. Moreover, in the special case of non-relativistic velocities, the BPS collision evolves adiabatically, meaning that the scalar field is always in a BPS antikink configuration, slowly evolving in moduli space with time, while at each instant the fermion field lies exactly at its respective zero mode.

We quantified fermion transfer between solitons through the computation of Bogoliubov coefficients similar to the ones studied in \cite{gibbons2007fermions,saffin2007particle}. After the collisions in most cases, the Bogoliubov coefficients reach a constant value which quantifies the probability that fermion is transferred from one state to the other. We found that close to the BPS case most fermions stay localized on the initial soliton except for high values of the coupling constant $g$. Moreover, as the initial velocity $v$ increases, the system moves further away from the BPS regime and more fermions are transferred to the other defect and to higher excited states or lost as radiation. On the other hand, for the non-BPS cases the fermions are much more likely, and in a higher amount, to be transferred to the other defect or excited states and the coefficients are more sensitive to the parameters of the model. 

An interesting continuation of our work can be to allow the defects to receive the fermion back-reaction. Thus, the soliton collisions should be altered as well as the soliton shapes. This would make the analysis based on the Bogoliubov coefficients less straightforward. However, we expect that some of our main results should be maintained. We plan to investigate this in a future work.

\appendix

\section{Numerical technique}
\label{ap1}

To integrate the field equations numerically we divide spacetime in a grid with spacings $\tau=h=0.01$. The scalar field at the gridpoint $(x_i,t_j)$ is $\phi_{i,j}$, where $i=0,1,\ldots,N$ and $j=0,1,\ldots,M$. Similar definitions are made for the fermion field. We approximate the spatial derivatives by a second order finite difference. For the scalar field we have, for instance,
\begin{equation}
\frac{\partial^2 \phi_{i,j}}{\partial x^2}=\frac{\phi_{i+1,j}-2\phi_{i,j}+\phi_{i-1,j}}{h^2}.
\end{equation}
The time integration is done using a Runge-Kutta fourth-order algorithm. The boundaries are set at $x=\pm 100.0$, giving $N=20000$. Boundary conditions are $\psi_{0,j}=\psi_{N,j}=0.0$ and $\phi_{0,j}=1.0$, while $\phi_{N,j}=\pm 1.0$, depending on the case considered. We evolve the system to a final time $t_f$ int the range  $100.0<t_f<400.0$, which is short enough so that the boundary conditions do not interfere with the bulk evolution. The fermion field is initially normalized to one and the time evolution using this method conserves the normalization with errors of the order $10^{-5}$ or less.

\section*{Acknowledgments}

We acknowledge financial support from the Brazilian agencies CAPES and CNPq. AM also thanks financial support from Universidade Federal de Pernambuco Edital Qualis A.


\end{document}